# Mobile Image Analysis Application for Mantoux Skin Test

Liong Gele[1] and Tan Chye Cheah[2]
[1]School of Computer Science, University of Nottingham Malaysia, Broga Road, 43500, Semenyih, Selangor, Malaysia.
[1*]lionggele40@gmail.com
[2]School of Computer Science, University of Nottingham Malaysia, Broga Road, 43500, Semenyih, Selangor, Malaysia.
[2]chyecheah.tan@nottingha.edu.my

* Corresponding Author

**Abstract:** This paper presents a newly developed mobile application designed to diagnose Latent Tuberculosis Infection (LTBI) using the Mantoux Skin Test (TST). Traditional TST methods often suffer from low follow-up return rates, patient discomfort, and subjective manual interpretation, particularly with the ball-point pen method, leading to misdiagnosis and delayed treatment. Moreover, previous developed mobile applications that used 3D reconstruction, this app utilizes scaling stickers as reference objects for induration measurement. This mobile application integrates advanced image processing technologies, including ARCore, and machine learning algorithms such as DeepLabv3 for robust image segmentation and precise measurement of skin indurations indicative of LTBI. The system employs an edge detection algorithm to enhance accuracy. The application was evaluated against standard clinical practices, demonstrating significant improvements in accuracy and reliability. This innovation is crucial for effective tuberculosis management, especially in resource-limited regions. By automating and standardizing TST evaluations, the application enhances the accessibility and efficiency of TB diagnostics. Future work will focus on refining machine learning models, optimizing measurement algorithms, expanding functionalities to include comprehensive patient data management, and enhancing ARCore's performance across various lighting conditions and operational settings.

## 1 Introduction

The escalating prevalence of Tuberculosis (TB), identified by the World Health Organization (WHO) as a leading cause of death from infectious diseases, poses a significant challenge to global health, particularly in achieving the United Nations' Sustainable Development Goal of health for all by 2030 [1]. TB is the leading killer of a single infectious agent worldwide, with 95% of cases and fatalities occurring in resource-poor settings [2-3]. In 2022, an estimated 10.6 million people fell ill with TB, with significant numbers in South-East Asia (46%), Africa (23%), and the Western Pacific (18%) according to the WHO's Global Tuberculosis Report 2023 [3]. The high incidence of TB in these regions is frequently linked to socioeconomic factors that exacerbate the disease's spread and impact.

Accurate diagnosis of Latent Tuberculosis Infection (LTBI) is crucial in managing TB, as LTBI can progress to active TB if left untreated. The Mantoux Tuberculin Skin Test (TST) is the predominant method for screening LTBI [4]. The Centres for Disease Control and Prevention (CDC) outlines a standard procedure for conducting a TST, involving the administration of tuberculin, measurement of the induration diameter using ballpoint pen and ruler method and interpretation of the results based on specific thresholds for different risk groups [4-6]. but it faces several challenges including errors in test administration, result interpretation, and patient follow-up compliance [5-6]. These issues are particularly problematic in resource-limited areas with inexperienced medical staff, leading to potential misdiagnoses and loss of diagnostic data [7-8].

To address these challenges, the Division of Biomedical Engineering at the University of Cape Town is developing mobile technologies for LTBI screening [9]. These prototypes aim to improve usability and diagnostic accuracy through a user interface for image capture and a backend for image analysis and result generation. The integration of mobile health (mHealth) technologies offers a promising solution to enhance the accuracy and comfort of the TST, enabling more precise and standardized methods for measuring induration, reducing errors, and improving patient adherence to testing schedules [9-11].

This paper proposes an mHealth tool to enhance the accuracy, standardization, and patient adherence in TST administration, improving patient care and outcomes in the fight against tuberculosis. The objectives are: (a) to develop image processing algorithms capable of identifying swelling areas, (b) to develop an algorithm for latent tuberculosis diagnoses based on the identified swelling areas, and (c) to integrate the developed image processing and machine learning model into a user-friendly mobile application.

This paper is organized as follows. Section 2 discusses the background and related work concerning the measurement of the diameter of Mantoux Skin Test. Sect. 3 describes the system development of design (methodology). Then, Sect. 4 describes the usability and real-time performance evaluation that we conducted along with the results. Finally, Sect. 6 concludes the paper and outlines future works.

## 2 Related Works

A systematic literature review following PRISMA 2020 guidelines was conducted to gather research on LTBI diagnostics through mhealth and skin lesion classification applications to identify the gaps and refine them. The search included databases like MDPI, PubMed, Google Scholar, IEEE Xplore, and others, focusing on studies from 2013 to 2023. Two search strings were used: one for TB diagnostics through TST and another for skin disease classification using machine learning. The first search resulted in 1,000 unique entries and 3 relevant studies: the second yielded 3,400 unique entries and 30 relevant studies.

## 2.1 The existing knowledge related to the mhealth of the diagnosis of LTBI in TST

The three pappers are from 2017 to 2020, research on LTBI diagnosis using the Tuberculin Skin Test (TST) advanced significantly [9-11]. In 2017, a study used a Samsung Galaxy S7 Edge for image capture and Agisoft PhotoScan for 3D reconstruction, involving stages such as finding common points in images, refining camera calibration, building a point cloud model, creating a polygonal mesh, and adding texture [10]. This method showed high intraclass correlation coefficients, indicating strong agreement with clinical methods but faced challenges like shadowing effects and limited resources. In 2018, further innovations included an automated induration measurement feature post-3D reconstruction, streamlining the workflow and surpassing manual methods. This comprehensive Android app for patient data collection emphasized user experience, recommending a more intuitive interface [9]. The study also aimed to improve screening throughput and reduce reliance on follow-up visits, but it highlighted the need for further real-world testing. By 2020, deep learning and Generative Adversarial Networks (GANs) were employed to create synthetic induration images, simplifying the analysis process and improving diagnostic accuracy [11]. However, challenges such as the need for larger and more diverse datasets and improved segmentation algorithms remain.

Several areas for improvement have been identified, including the necessity for larger datasets to train deep learning models, alternative image assessment methods, and enhancements in usability and diagnostic precision. While GANs show promise, thorough validation is required. Extensive real-world testing focused on cost-effectiveness and accessibility, especially in resource-constrained settings, is essential. Revising depth map segmentation methods and developing machine learning-based approaches have been recommended for detecting and segmenting indurations, eliminating the need for scaling stickers.

## 2.2 The existing knowledge related to the mhealth for classifying Skin Lesion

The evolution of mobile applications in dermatology, particularly for real-time skin lesion analysis, has advanced significantly through key studies. These studies, featuring unique approaches and technological innovations, have greatly contributed to skin cancer detection and diagnostics by utilizing advanced image processing, machine learning, and deep learning techniques.

Study [12] proposed an augmented reality mobile application for dermatologists, capturing live video snapshots of lesions, tracking device position, and zooming in for detailed analysis. It uses a Convolutional Neural Network (CNN) to assess signs of skin cancer based on symmetry, border, color, diameter, and texture, displaying enhanced images for comprehensive real-time analysis. The CNN model, trained on a desktop PC and uploaded to an Android smartphone, leverages TensorFlow. Study [13] demonstrated a real-time skin lesion detection method on Android devices using the pre-trained SSD-Mobilenet model from TensorFlow, allowing for swift and accurate lesion detection and localization. Study [14] focuses on real-time diagnosis using a mobile app powered by CNNs and the HAM10000 dataset, capturing high-resolution images that are pre-processed and analysed in real-time. The integration of TensorFlow and Keras, optimized through TensorFlow Lite, enables prompt and accurate diagnosis. Other notable applications include DermEngine [15], which monitors nevus evolution by associating nevus pictures with specific body positions, and SkinVision [16-17], a smartphone app that captures lesion images and provides risk assessment based on grey-scale image analysis, achieving a sensitivity of 97% and specificity of 78%.

These studies collectively embody cutting-edge mobile application development in dermatology, each offering a unique real-time processing approach and device-specific adaptation to skin lesion analysis. Moreover, mobile technology significantly advances lesion detection by blending real-time processing with advanced imaging and augmented reality (AR). For instance, Study [12] highlights AR's use in dermatological assessments, enhancing the visualization of critical diagnostic factors like asymmetry and border irregularities. ARCore's sophisticated environmental understanding and Depth API create precise depth maps, enabling accurate lesion localization. Several studies align with our research objectives, particularly in developing algorithms for latent tuberculosis diagnoses. Advanced deep learning architectures like SAM [18], YOLOv4 [19], and DeepLabV3+ [20] have proven effective in medical image analysis and are suitable for our project. These models, supported by resources detailing their adaptation to mobile platforms, promise improved diagnostic accuracy and usability, especially in resource-constrained environments, significantly contributing to global health technology.

## 3 System Development and Design

### 3.1 Methodology

The methodology for developing a mobile-based solution for Tuberculin Skin Test (TST) evaluation begins with requirement gathering through stakeholder interviews with TB experts to understand diagnostic procedures and the potential of mobile apps in TB management. A systematic literature review adhering to the PRISMA 2020 framework identified gaps in current LTBI diagnosis methods. Design criteria were then established. The project leverages AR technology (ARCore) for depth estimation in 2D images [21], using Android Studio's CameraX API for efficient image processing [22]. A diverse dataset of annotated skin images was curated, and various machine learning models, including SAM, Yolo Instance Segmentation, and DeepLabV3, were developed and evaluated using metrics like accuracy, precision, recall, and F1 score. These models were integrated into an Android application for real-time TB diagnosis, incorporating user feedback to continuously improve performance. The system approach was finalized, outlining technical architecture, followed by UI/UX design based on atomic design methodology [23]. Iterative prototype development and testing were conducted to ensure reliability and usability. Finally, system evaluation in real-world scenarios assessed the application's performance, with usability testing methods like Level of Agreement providing valuable user feedback [9, 24].

### 3.2 System Development and Design

### 3.3 The proposed System

The system we propose supports clinicians during the Tuberculin Skin Test (TST) by providing a high-level overview of real-time skin induration analysis using a mobile application. The system's architecture follows a structured sequence, starting with the mobile unit, which captures skin images in real-time. The application interface employs continuous tracking to ensure the induration remains in precise focus, while Augmented Reality (AR) technology assesses depth values, providing a three-dimensional perspective for accurate appraisal. Orientation sensors correct the device's positioning to mitigate involuntary movement and enhance capture stability, maintaining image capture integrity. The user-centric design includes an auto-capture feature that automatically takes the image when the induration is correctly aligned, enhancing the user experience by eliminating manual input. Post-capture, images are securely stored and undergo pre-processing, including noise reduction and contrast enhancement, to ensure optimal quality for analysis. Image segmentation is conducted, extracting key features such as edges and color variations for classification. A machine learning component precisely measures the induration's diameter, ensuring accuracy in alignment with actual physical dimensions. The system's operation culminates in comprehensive image analysis and classification, categorizing the induration and indicating a positive or negative TB test result. This includes an in-depth examination of the extracted features and measurements. The mobile unit facilitates patient information retrieval and image acquisition, capturing necessary images for storage and processing. The evaluation is completed as the measurement and analysis phase takes place, representing the final steps in this innovative process.

## 4 Real-time processing

### 4.1 Considerations for Clinical Evaluation of TST

When assessing the Tuberculin Skin Test (TST), guidelines dictate measuring the induration width perpendicular to the arm's longitudinal axis. This ensures proper alignment and clear demarcation of the induration's boundaries for precise measurements. To train for TST induration size recognition, modelling clay indurations (5mm, 10mm, and 15mm) were created to mimic the raised skin reaction of a positive TST result. Accurate measurements were obtained by considering factors such as optimal lighting, camera angle, and depth perception.

### 4.2 Image Processing Algorithm

To display information in real-time on the user's camera, TBCheck requires continuous tracking to monitor the patient's skin induration as the clinician moves the device and determines the distance from the user's skin. The application suggests enabling camera access for AR functionality, offering user permissions. It prompts users to activate the depth-sensing feature supported by ARCore, enhancing realism by detecting object geometry in real space. ARCore is used for skin surface detection and depth analysis, incorporating motion tracking and environmental understanding. During motion tracking, the camera detects flat skin surfaces and uses Concurrent Odometry and Mapping (COM) to determine its position and orientation based on feature point detection and IMU data. In the environmental understanding stage, feature points from the skin surface are aggregated into planes, rendered with materials and shaders to conform to the skin's contours, and continuously updated as the camera moves to ensure accurate tracking. The Depth API generates 3D depth images, allowing precise placement of virtual objects and handling interactions like occlusion. The *getMillimetersDepth* function extracts precise measurements down to the millimeters, ensuring accurate spatial representation. Practical implementation involves capturing a broader array of feature points for defining planes and understanding depth, with optimization techniques like using a contrasting background to enhance plane detection.

### 4.3 User Position and Image Capture Protocol

Advanced sensor-based orientation handling tracks and visualizes orientation changes in real-time, ensuring stable image capture. The application implements protocols for capturing induration images using Depth Value which obtain with the help of ARCore and Orientation Panels, guiding users to optimal settings before automatic image capture. Users are prompted to hold the device correctly, with the palm behind the phone and the thumb on the volume button for stability. The image capture protocol includes using the Depth Value Panel (175-400 mm) and Orientation Panel (0-degree pitch and roll) to ensure accurate image acquisition. The distance between the device and the patient skin is measured through experiment. An induration guide with green and blue outline circles assists in positioning the induration within the correct area. The Auto Capture function, implemented through the CameraX API, automatically captures images once all protocols are met, eliminating manual input and ensuring uniformity. The captured images are cropped to a center region of 450 x 450 pixels for clarity and stored in Firebase for analysis. The app allows users to review the captured image and provides options to retake it or view analysis results.

### 4.4 Machine Learning Image Segmentation

Testing of segmentation models, including YOLO Instance Segmentation, SAM, and DeepLab v3, led to the successful implementation of DeepLab v3 for accurate induration analysis. A dataset of 2,298 annotated skin lesion images from PAD-UFES-20 [70] was used for training, ensuring precise delineation of TB indurations. The best-performing model, trained with DeepLab v3 using a ResNet-50 backbone, was converted into a mobile-friendly version via PyTorch's optimization techniques, facilitating real-time usage on mobile platforms. The application loads the optimized model (*.ptl* file), converts images into tensors, and normalizes them for consistent data. The model outputs a segmentation map, categorizing each pixel based on features like induration. This map is overlaid on the original image in two versions: one semi-transparent and one opaque, enabling efficient segmentation directly on the device. After capturing an image, users can review and retake it if necessary. Upon approval, the segmentation and analysis process generate a detailed map of the skin induration. Successful detection depends on good lighting and distinct color contrast, while poor conditions can lead to inaccuracies. Segmentation results showed a steady increase in training accuracy and a decrease in validation loss, demonstrating the model's ability to generalize. The model was integrated into the mobile application, ensuring reliable and accurate induration analysis.

### 4.5 Automated Measurement of Induration

The app measures induration diameters using the Euclidean Distance Formula $d = \sqrt{(x_2 - x_1)^2 + (y_2 - y_1)}$ converting pixel measurements into millimeters for accurate tracking. It identifies the induration, highlights its edges, and determines the longest line that can be drawn across the highlighted edge. This distance, initially measured in pixels, is then converted into *millimeters* using a calibrated conversion factor that varies dynamically with the measured pixel diameter: 0.1197 if the max diameter is less than 50 pixels, 0.1523 if the max diameter is between 50 and 80 pixels, and 0.1499 if the max diameter is between 80 and 200 pixels. The app employs edge point detection to traverse each pixel, identify matches with the segment color, and confirm edge use via neighbouring pixels. This functionality is vital for tracking changes over time and essential for effective treatment and monitoring, as validated in Experiment 4: Skin Induration Experiment + Evaluation.

### 4.5 Patient Information Collection and Reminder Function

This mobile application collects patient-specific information, including health condition and exposure to TB, to determine the induration size range for a positive LTBI result. A questionnaire mimics the questions used to interpret the induration size range. The reminder function helps maintain user engagement and adherence to testing protocols, ensuring timely follow-ups. The interface allows users to set up reminders for their TB skin tests, providing easy navigation through tutorial information and test reminders. Real-time notifications ensure timely documentation and follow-up, integral to accurate monitoring and results analysis.

## 4.6 Evaluation

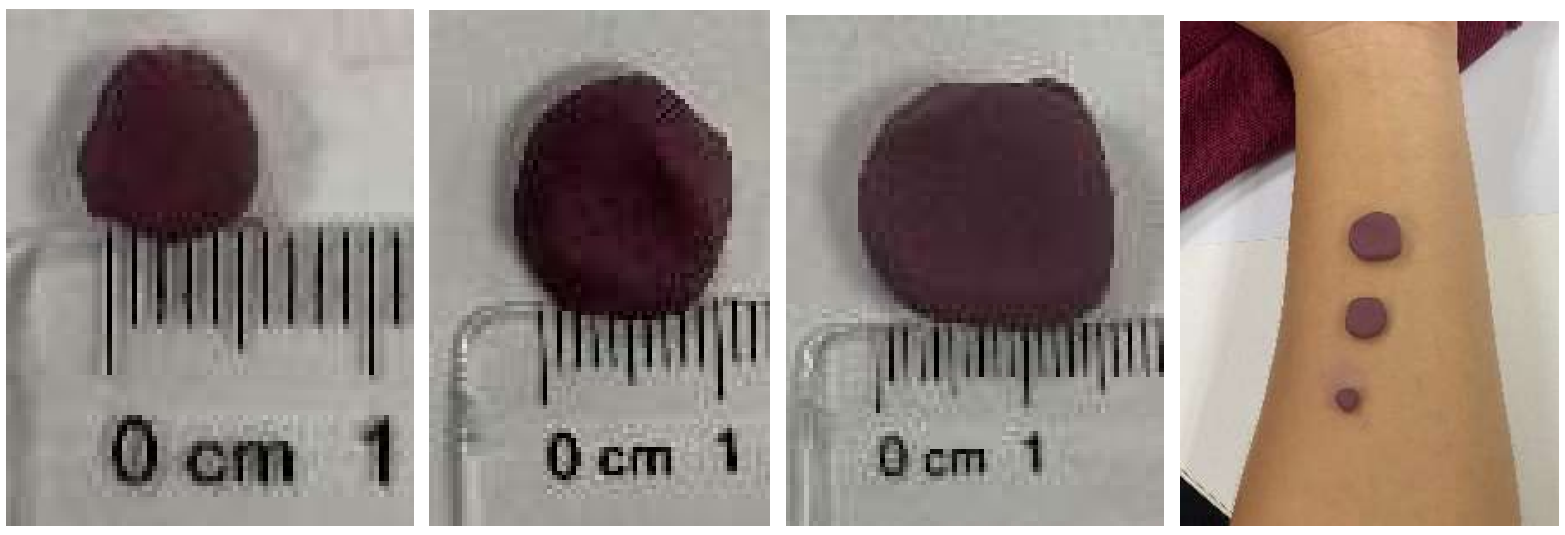

*Figure 1: Size Comparison of Modeled TB Indurations (5mm, 10mm, 15mm) Using Clay with Ruler for Scale. Figure 3: Demonstration of Modeling Clays Simulating TB Indurations on a Forearm for Diagnostic Training.*

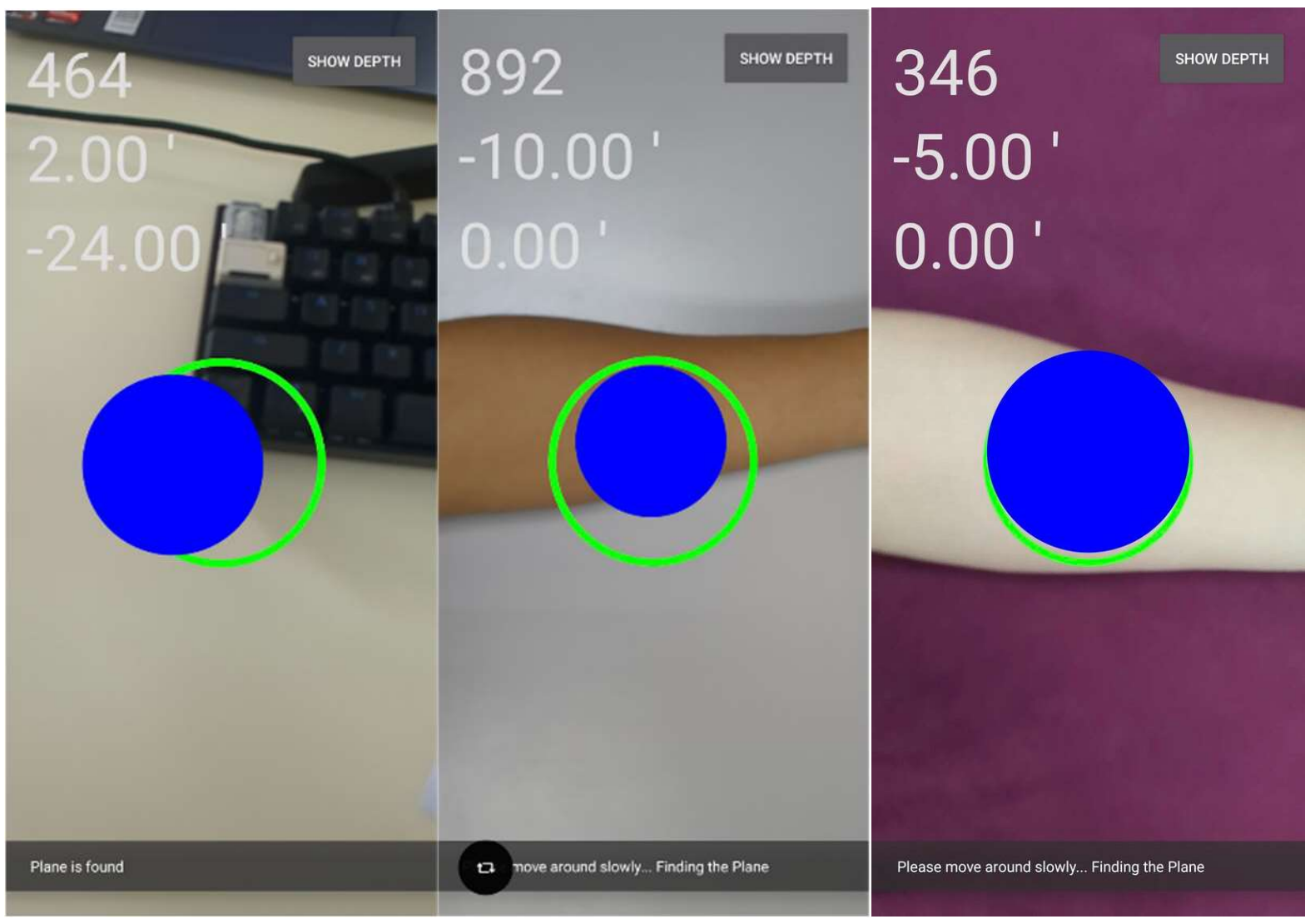

*Figure 2: Comparison of Plane Detection Without Cloth (Middle) and With Cloth (Right).*

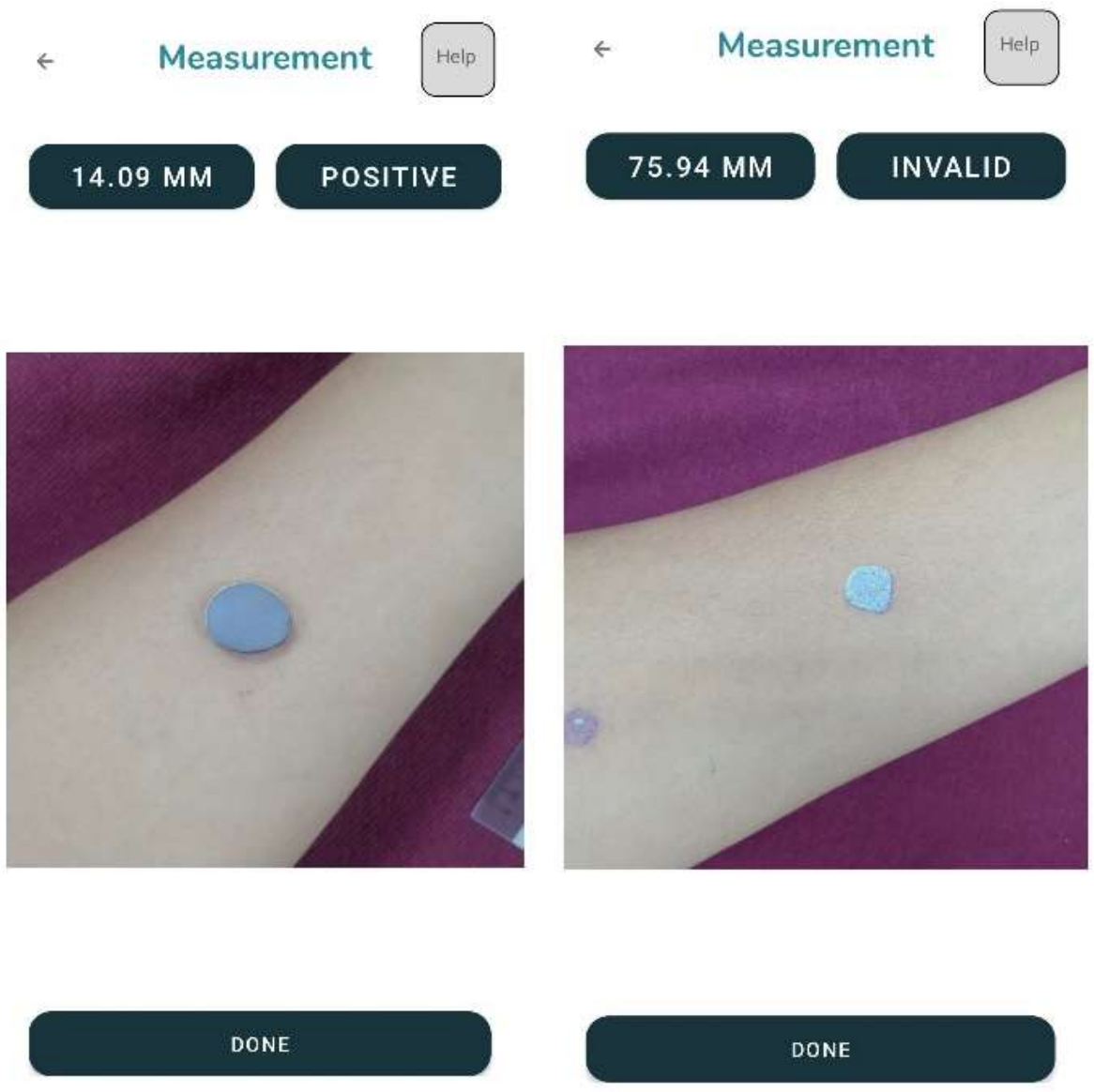

*Figure 3: Successful and Failed Segmented of Skin Induration*

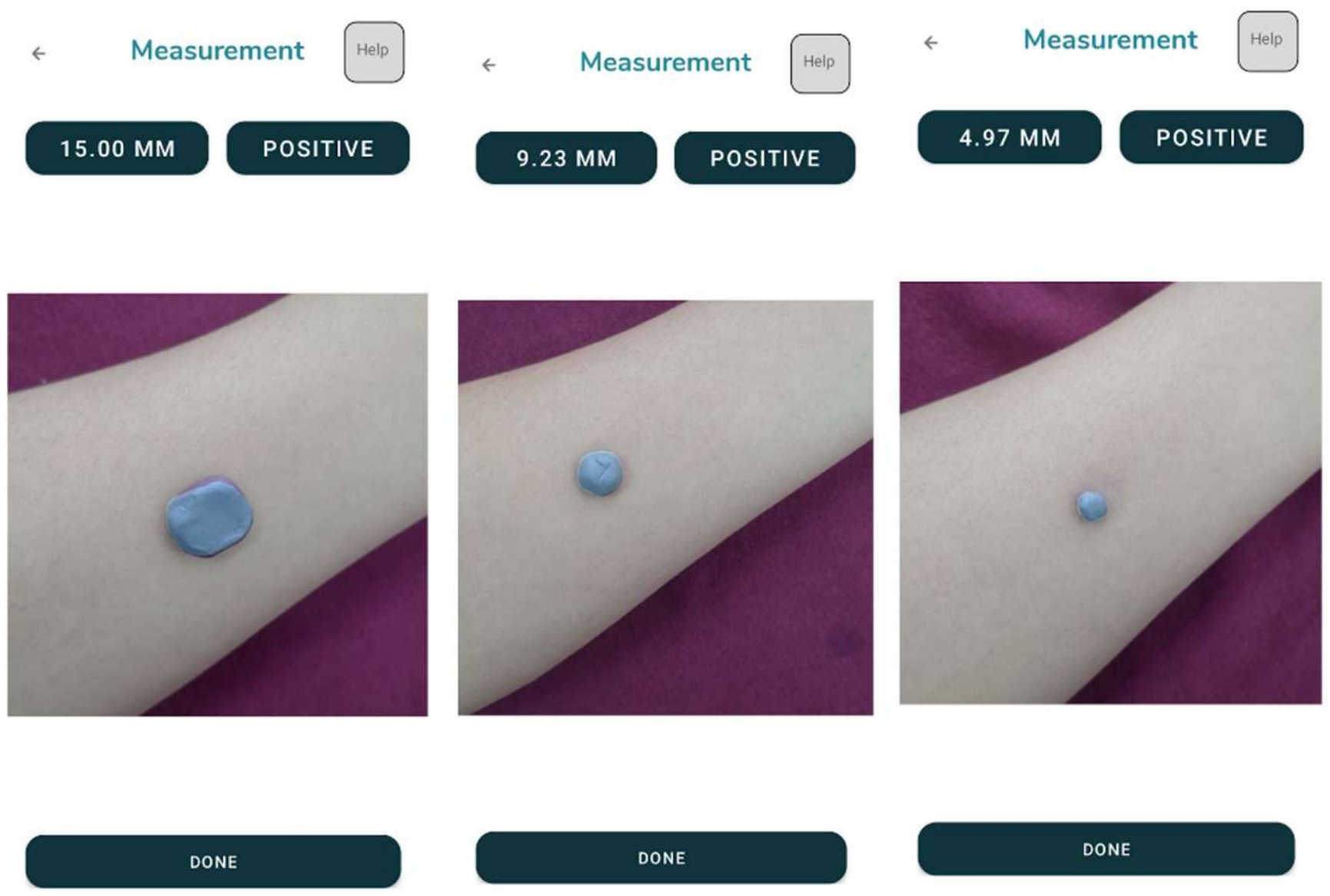

*Figure 4: Induration Measurements of 15.00 mm (left), 9.23 mm (middle), and 4.97 mm (right). Positive result when induration diameter is 10-15 mm and the checkbox 'Have you lived in a low incidence TB area?' is selected.*

Experiments were conducted to enhance the mock of Tuberculin Skin Test (TST) analysis and the development of the application and system using mobile application. Experiment 1 evaluated machine learning models—SAM, YOLO Instance Segmentation, and DeepLabV3—using the PAD-UFES-20 dataset. Divided into training, validation, and testing sets with data augmentation, YOLO showed promise but struggled in mobile settings, while SAM was effective in controlled environments but faced integration challenges. DeepLabV3 excelled in real-time analysis, making it the best choice for the mobile application. Experiment 2 focused on determining the optimal depth value for accurate TST induration measurement. Images of a 10mm induration were captured at various depths, with the optimal range for accuracy being 219-220mm, where the predicted measurement was 9.91mm, demonstrating precise and reliable results in clinical settings. Experiment 3 sought to find the optimal scalar factor for TST induration measurement using mock indurations of 5mm, 10mm, and 15mm. Optimal scalar factors were 0.1197 for 5mm, 0.1523 for 10mm, and 0.1399 for 15mm. The higher variability in larger indurations suggests further fine-tuning is needed for accuracy. Experiment 4 assessed usability with ten participants measuring mock indurations and completing a level of agreement. While 60% felt confident using the app, only 50% found it easy to use, indicating a need for simplification and better guidance. Despite 74% appreciating the user experience, 80% needed assistance, highlighting interface challenges and suggesting improvements in layout and navigation.

## 5 Conclusion and Future Works

The study successfully demonstrates the potential of integrating advanced image processing and an algorithm for latent tuberculosis diagnoses based the identified swelling areas with the machine learning technologies into mobile health applications to improve the diagnostics of LTBI. Through systematic research and development, the project highlights significant advancements in mobile technology that can be applied to tuberculosis screening. The application developed not only enhances the accuracy of TST but also offers a more accessible and efficient approach to tuberculosis diagnosis in diverse healthcare settings. Future work should focus on refining these technologies, expanding their applicability, and validating their effectiveness in clinical environments to ensure they meet global health standards and contribute meaningfully to the fight against tuberculosis.

Future work should focus on refining DeepLabv3 models to improve image segmentation accuracy, particularly in isolating indurations from dermatoscopic images and accommodating variations in skin types. Implementing SAM models could further enhance segmentation precision. Optimizing diameter measurement algorithms and scalar factors by leveraging ARCore's depth sensing capabilities is essential. Additionally, developing a comprehensive patient data management system with secure data protocols is crucial. Enhancing ARCore's stability and accuracy under varied lighting conditions, possibly through hybrid approaches, should be explored. Addressing ethical and clinical testing requirements by obtaining approval and accessing real patient data will ensure clinical applicability. Expanding training datasets to include diverse demographics will help train robust models. Lastly, improving the user interface for image capture and analysis to ensure intuitiveness and accessibility, especially in resource-limited settings, is vital for successful implementation.